\documentclass[runningheads]{svmult}

\usepackage{makeidx}   
\usepackage{graphicx}  
\usepackage{subeqnar}  
\usepackage{multicol}  
\usepackage{physprbb}  
\makeindex             

\begin{document}
\title*{Super-massive stars: Dense star-gas systems}
\toctitle{Super-massive stars: Dense star-gas systems}
%
%
\titlerunning{{$\cal SMS$}s: Dense star-gas systems}
%
\author{Pau Amaro-Seoane
\and Rainer Spurzem
\and Andreas Just}
\authorrunning{Pau Amaro-Seoane et al.}
%
%
\institute{Astronomisches Rechen-Institut, 
M\"onchhofstra\ss e 12-14\\
D-69120 Heidelberg, Germany\\
pau, spurzem, just@ari.uni-heidelberg.de}

\maketitle              

\begin{abstract}
We use a gaseous model and a semi-analytical approach 
to study the evolution of a super-massive central gaseous
object (a super-massive star, {$\cal SMS$} from now on) in
an AGN and its evolution by interactions with the surrounding stellar
system. Our future work in this field is outlined, which aims at a
more detailed study
of energy flows in the interstellar medium, stellar evolution and the
relation between QSOs and galaxy formation.
\end{abstract}

\subsection*{Super-massive Stars, Galaxies and Cosmology}
Several theoretical models have been proposed in order to explain
the properties of quasars and other types of active galactic nuclei 
(hereafter AGNs). In the 60s and 70s super-massive central objects 
(from now onwards SMOs) were thought to be the main source of their 
characteristics. {$\cal SMS$}s and super-massive black holes (SMBHs) are two 
possibilities to explain the nature of these SMOs being harboured in the 
AGNs. These super-massive gaseous objects may be the precursors of the SMBHs
and those may be an intermediate step towards the formation 
of these. Large amounts of gas lost by stars during their evolution will stock in 
galactic centres.
In previous work (Amaro-Seoane \& Spurzem, 2001) we analyze
such a physical system in spherical symmetry.
We aim to carry out in the 
next time the study of the
evolution of the {$\cal SMS$} and the stellar system by performing a 
more detailed semi-analytical
evaluation of the interaction between stars and gas
(cf. Deiss, Just, \& Kegel 1990), examining the global 
stability of the central object, and its dependence on the loss-cone stars 
that heat it and determine its contraction rate.
A detailed numerical model is the via 
to do it. 

Nowadays there is strong evidence that the formation of central
black holes in galaxies can be qualitatively understood in the
framework of cosmological hierarchical galaxy formation (Haehnelt \& Kauffmann
2000, Kauffmann \& Haehnelt 2000, see also earlier work of e.g. Eisenstein
\& Loeb 1995). We stress that, although the embedding into cosmological
scenarios had advanced very much during the past two or three decades since
the times of Spitzer, Colgate and others, the detailed study of star-gas 
interactions in dense nuclei, is still incomplete and very important for an
understanding of the physical processes at work.

\subsection*{Core Structure and Evolution of ${\cal SMS}$s}

Hoyle \& Fowler (1963) dropped already the hint that maybe at the centres of 
galaxies star-like objects exist
with masses of up to $10^8M_{\odot}$. 
As regards the stability and/or existence of such an object, they ``turned a 
blind eye''. They developed their 
suggestion with the argument that wholly convective stars could ``do the job''. 
When radiation pressure is dominant, the convective condition is expressed by 
a polytrope of index three. The polytropic index is very near to the stability limit 
and post-Newtonian corrections lead to an instability before the 
central temperature allows nuclear burning for masses
\begin{equation}
{\cal M} > 3\cdot 10^5 M_{\odot};\ \ 
{\cal R}_{\rm crit} = 0.086 {\rm pc} \cdot {\cal M}_8^{3/2}
\end{equation}
\noindent
where ${\cal M}$ is the mass of the {$\cal SMS$} and
${\cal R}_{\rm crit}$ is the corresponding
critical radius, where the instability sets in
(${\cal M}_8$ is the mass in units of $10^8M_{\odot}$). 
Therefore, an isolated {$\cal SMS$}
eventually run into a catastrophic collapse. The outcome (super-massive black
hole, re-bounce, disruption of the {$\cal SMS$}, ...) depends very much on the details of
its history. Thus, the study of not just the time evolution of the system,
but of its stability and, thence, of the general dimensions of the central 
object seems to be of crucial importance to get a rigorous understanding of the 
whole problem.
Energy and mass transport processes, among them so-called ``energy diffusion'' 
and trapping of loss-cone stars
(Frank \& Rees, 1976, Hara 1978) control the global polytropic structure
of the $\cal SMS$ and its evolution. They strongly depend
on a correct physical description of the interactions between
the gaseous and stellar components by, for instance, dynamical friction. 

Following Hara (1978) and Langbein et al. (1990) 
we select two competing star-gas 
interaction processes in our dense gas-star system. The first one is
the contraction (or collapse with re-bounce) of the mixed core with 
stars that are
trapped by friction within the gas and slow down additionally to build a new highly condensed
stellar core. This core may re-heat the gas or decouple to build a new core.
Second,
a core-halo interaction is considered, where loss-cone stars of the surrounding 
stellar
system heat the core and feed it by means of new trapped stars. This core collapse can be conked
for a while until the loss-cone is ``empty'' (not rigorously speaking,
but for the practice we can assume it is empty) or the
core becomes too massive. The result may be a sequence of {$\cal SMS$}-stars cores contained one
inside the other until the relatively low mass innermost {$\cal SMS$} initiates hydrogen burning or
collapses to a BH ``seed''. In Fig. 1 we give an intuitive scheme for this 
scenario.

\begin{figure}[h!]
\begin{center}
\includegraphics[width=.6\textwidth]{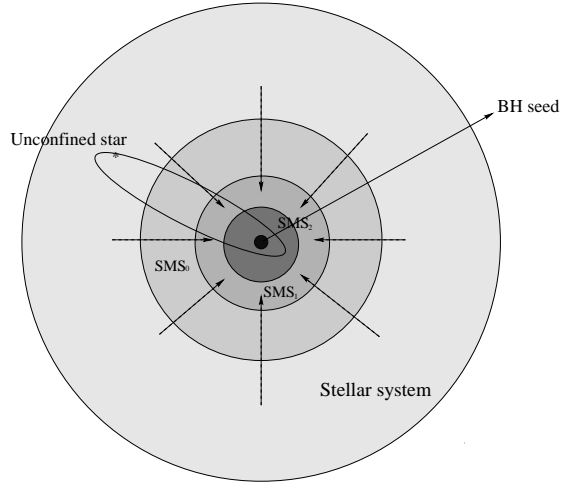}
\end{center}
\caption[]{Sequence of {$\cal SMS$}-stars cores: {$\cal SMS$$_{2}$}$\subset$ 
{$\cal SMS$$_{1}$}$\subset${$\cal SMS$$_{0}$}. The non-confined stars which belong to
the loss-cone sink onto the central object after a number of crossings, which we call
the ``trap'' number. The innermost {$\cal SMS$} may collapse to a BH seed.}
\label{eps1}
\end{figure}

\subsection*{The gaseous model}

To go from stationary to dynamical models we use a gaseous model of
star clusters in its anisotropic version (see e.g. Louis \& Spurzem 1991,
Spurzem 1994, Giersz \& Spurzem 1994, and also {\tt http://www.gaseous-model.de}
for the more recent developments). The basic idea is that of a 
kinetic equation of the Boltzmann type with the inclusion of a 
self-consistent collisional term of the Fokker-Planck type. 
By taking velocity moments of such Fokker-Planck equation of up to order two
and closing the system with a heat flux closure we get a set of moment
equations which is very similar to gas-dynamical equations and it is then
coupled to Poisson's equation for the total gravitational potential.
Such a model is very well
suitable
to treat collisional and collisionless evolution of coupled stellar and star-gas
systems, because the stellar dynamical equations resemble gas dynamical ones.

\subsection*{Acknowledgements} This work has been performed in SFB439 ``Galaxies
in the Young Universe'' at the Univ. of Heidelberg. We are grateful to S.
Deiters for providing the gaseous model web page.

%

\end{document}